\makeatletter
\let\saved@PackageWarningNoLine\PackageWarningNoLine
\def\PackageWarningNoLine#1#2{%
  \def\@tempa{#1}\def\@tempb{inputenc}%
  \ifx\@tempa\@tempb\else
    \saved@PackageWarningNoLine{#1}{#2}%
  \fi
}
\makeatother

\documentclass{aa}  
\makeatletter
\let\PackageWarningNoLine\saved@PackageWarningNoLine
\makeatother
\usepackage{graphicx}
\usepackage{txfonts}

\makeatletter
\AtBeginDocument{%
  \@ifpackageloaded{hyperref}{%
    \def\Hy@Warning#1{}%
  }{}%
}
\makeatother
\usepackage{hyperref}

\makeatletter
\let\saved@PackageWarning\PackageWarning
\def\PackageWarning#1#2{%
  \def\@tempa{#1}\def\@tempb{caption}%
  \ifx\@tempa\@tempb\else
    \saved@PackageWarning{#1}{#2}%
  \fi
}
\makeatother
\makeatletter
\let\PackageWarning\saved@PackageWarning
\makeatother
\usepackage{amsmath,amssymb}
\usepackage[T1]{fontenc}

\newcommand{\diff}{\mbox{${\rm d}$}}
\newcommand{\av}{\mbox{$A_V$}}
\newcommand{\mh}{\mbox{\rm [{\rm M}/{\rm H}]}}
\newcommand{\Msun}{\mbox{$M_{\odot}$}}
\newcommand{\Rsun}{\mbox{$R_{\odot}$}}
\newcommand{\Teff}{\mbox{$T_{\rm eff}$}}
\newcommand{\logT}{\mbox{$\log\Teff$}}
\newcommand{\logte}{\mbox{$\log\Teff$}}
\newcommand{\logg}{\mbox{$\log g$}}

\newcommand{\BC}{{\rm BC}}
\newcommand{\teff}{T_{\rm eff}}
\newcommand{\logteff}{{\rm log}\,T_{\rm eff}}

\makeatletter
\AtBeginDocument{%
  \@ifpackageloaded{natbib}{%
    \renewcommand\bibcite[2]{%
      \global\@namedef{b@#1\@extra@binfo}{#2}%
    }%
  }{}%
}
\makeatother

\newcommand*\samethanks[1][\value{footnote}]{\footnotemark[#1]}

\begin{document}

\title{Bolometric correction for cosmologically redshifted stars with dust:}

\subtitle{an update to the YBC database}

   \author{Yang Chen \inst{1}\fnmsep\thanks{Corresponding authors}
          \and
           Xiaoting Fu \inst{2}\fnmsep\samethanks
           \and
           L\'eo Girardi \inst{3}
           \and
           Helge Todt \inst{4}
           \and
           Alessandro Bressan \inst{2,3,5}
          }

   \institute{School of Physics, Anhui University, Hefei, 230601, China\\
              \email{cy@ahu.edu.cn}
         \and
             Purple Mountain Observatory, Chinese Academy of Sciences, Nanjing 210023, China\\
             \email{xiaoting.fu@pmo.ac.cn}
             \and
             INAF - Osservatorio Astronomico di Padova, Vicolo dell'Osservatorio 5, Padova, Italy
             \and
             Institut f\"{u}r Physik und Astronomie, Universit\"{a}t Potsdam, Karl-Liebknecht-Str. 24/25, 14476 Potsdam, Germany
             \and
             SISSA, via Bonomea 365, I--34136 Trieste, Italy
             }

\date{Received ; accepted }

\abstract{
Observations from Hubble Space Telescope (HST) and James Webb Space Telescope (JWST) continue to reveal gravitationally magnified high-redshift star candidates, resulting in an increasing demand for accurate stellar bolometric corrections to facilitate comparisons between theoretical stellar evolutionary models and observational data. To address this need, we have updated our YBC stellar bolometric correction database by incorporating bolometric corrections for cosmologically redshifted stars, which is called the zYBC database. These bolometric corrections are derived by redshifting stellar spectra from public stellar spectral libraries, attenuated by the extinction curves for both the dust in the host galaxy and the Milky Way, and followed by convolution with the transmission curves of various photometric filters. Our methodology naturally incorporates the effects due to the cosmological $K$-correction and the dust. In addition to the existing stellar spectral libraries in the previous version of YBC, we include stellar spectral libraries from non-local thermodynamic equilibrium (non-LTE) models (PoWR, TLUSTY and CMFGEN) for O, B stars, which are more appropriate for hot massive stars. Especially, the PoWR and CMFGEN models include the modelling of stellar wind. The database supports key photometric systems for high-redshift studies, such as HST/WFC3 wide-band filters, JWST/NIRCam, and Chinese Space-station Survey Telescope (CSST) equipped with its Main Survey Camera (MSC) and Multi-Channel Imager (MCI), which are pivotal facilities for the current and future observational studies of high-redshift stars, and maintains the flexibility to incorporate additional photometric systems upon request. As examples, we present colours as functions of $\logte$ at various redshifts for several photometric systems, which exhibit non-monotonic behaviours and demonstrate the necessity for a dedicated modelling. In particular, we find that the relations show larger dispersions at high redshift than the zero redshift case. This indicates that the stellar parameters of high redshift stars can be better determined than those of their local counterparts, given that their redshifts can be reliably determined through other methods (e.g., from their host galaxies) and their photometric data are of high enough quality for physical parameter determination through spectral-fitting method. We also show the difference in the effect brought by the different amount of extinctions. This updated database represents a valuable resource for high-redshift star research.}


\maketitle

\section{Introduction}

Accurate modelling of high-redshift stellar populations is essential for advancing our understanding of the first stars, cosmic re-ionisation, cosmic metal enrichment, cosmic dust evolution, galaxy formation and evolution, and stellar evolution in the early universe \citep{Madau2014,Smith2014,Vink2022,Klessen2023}. Stellar population synthesis (SPS) emerges as nearly the sole approach available for modelling these distant stellar populations \citep{BC03,sedfitting,Conroy2013}. This method relies heavily on stellar evolutionary tracks and spectral libraries, which are typically calibrated using observations from our local universe. However, it is important to recognise that the assumptions underlying these models may not fully apply to high-redshift stars, where conditions can differ significantly. This can only be tested with direct comparisons between stellar models and high-redshift star observations. The detection of Icarus, a high-redshift star candidate observed through gravitational lensing \citep{Rodney2018,Kelly2018}, marked a new era in directly studying the characteristics and behaviour of these ancient stellar objects. Subsequent studies have then revealed different populations at different redshifts, with the highest redshift one being at $z\sim$6.2 \citep{Welch2022a}. On-going observations with the Hubble Space Telescope (HST) and James Webb Space Telescope (JWST) continue to reveal gravitationally magnified high-redshift star candidates, and the forthcoming Chinese Space-station Survey Telescope (CSST), with its Main Survey Camera (CSST/MSC) and Multi-Channel Imager (CSST/MCI), is expected to yield further detections.

To derive the physical properties of stars from photometric data using stellar evolutionary models or to compare these models with observational data, it is necessary to establish a relationship between theoretical stellar parameters (e.g., mass, age, and composition) and observable photometric measurements. This connection can be established through the application of stellar bolometric corrections \citep{Bessell1998, Girardi2002, Casagrande2014, Chen2019}, which bridge the theoretical predictions with the observations.

Compared to the bolometric corrections for local observations, the cosmological $K$-correction is the major effect to be applied to the bolometric correction for high-redshift stars. The cosmological $K$-correction in a certain passband is defined as the corrective term to be applied to the observed photometry due to the spectrum being redshifted. This term can be positive, zero, or negative, depending on the interplay among the redshift, the spectral shape, and the passband. 

The cosmological $K$-correction is extensively studied for high redshift galaxies, supernovae or star clusters\citep{Kim1996,Blanton2007,Reina-Campos2024}, while for normal stars it is still largely untouched in a systematic way. For example, in \cite{Kelly2018} and \cite{Chenwenlei2019}, the $K$-correction used is from \cite{Kim1996} for type Ia supernovae rather than for normal stars. We therefore started to establish a publicly available database for the bolometric corrections of high-redshift stars with popular spectral libraries. Since most of the stars observed are massive stars, this work focuses on the bolometric correction for high-redshift massive stars.

For establishing the bolometric correction for high-redshift stars, one major difficulty is the lack of comprehensive spectral libraries of different metallicities for massive stars with the effect of stellar wind and non-local thermodynamic equilibrium (non-LTE) included. Regarding this aspect, spectral libraries computed with PoWR and WM-basic provide large grids of non-LTE models with different mass-loss rates \citep{PoWR2019,Chen2015}.

Another complication of computing the bolometric correction for high-redshift stars is dealing with the extinction. This includes extinction from the circumstellar dust, galactic dust of the host galaxy, intervening galactic dust, intergalactic medium (IGM), and the Galactic dust. Handling the circumstellar dust, galactic dust of the host galaxy and the Galactic dust is relatively trivial since these factors are affecting at fixed redshifts. While for the intervening galactic dust and IGM, they are affecting at random redshifts and cannot be easily parameterised. Therefore, in the present work, we only take into account the effect by the former but leave the latter for the future more specific work. 

This paper is organised as follows. We introduce necessary definitions and derive the formula for calculating the bolometric correction (BC) for cosmologically redshifted stars in section~\ref{sec:eq}. In section~\ref{sec:spec}, we introduce the stellar spectral libraries. In section~\ref{sec:bc}, we present the computed BCs and $K$-corrections. In section~\ref{sec:track}, we show the evolutionary tracks of redshifted stars on the colour-magnitude diagram with the BCs computed in this work. In section~\ref{sec:discuss}, we discuss the results.

\section{The formulations for computing the bolometric correction}
\label{sec:eq}
In this section, we derive the formula for calculating the bolometric corrections for cosmologically redshifted stars. Before doing so, we need to recall some basic relations concerning the bolometric correction for the zero-redshift case, which are necessary for the discussion, though \citet{Kurucz1979}, \citet{Bessell1998}, \citet{Girardi2002} and \cite{Chen2019} have already provided an exhaustive discussion on this topic. For simplicity, we only discuss the quantities for the present-day photon-counting systems, which are adopted by most of the modern photometric devices.

\subsection{Zero redshift case}
We maintain the notations as in \cite{Chen2019}. Most formulas for the zero redshift case were already given there, but we rewrite them here, both for better readability and as a reference for the non-zero redshift case later.

When the telescope receives a radiation flux $f_\lambda$ (or $f_\nu$) from a distant star, the magnitude in a certain passband $i$ with a transmission curve $S_{\lambda,i}$, delimited by the lower wavelength $\lambda_1$ and the upper wavelength $\lambda_2$, is defined as
\begin{equation}
  m_i \equiv -2.5{\rm log}\left[\frac{\int_{\lambda_1}^{\lambda_2} \lambda f_{\lambda} S_{\lambda, i} \diff\lambda}
    {\int_{\lambda_1}^{\lambda_2} \lambda f^0_\lambda S_{\lambda, i} \diff\lambda}\right] + m_{i,0}.
    \label{eq_mag}
\end{equation}
In this equation, $f^0_\lambda$ is the flux of the reference spectrum and $m_{i,0}$ is the corresponding reference magnitude. Depending on the reference spectrum used and the reference magnitude defined, different magnitude systems are defined, among which the Vega and AB magnitude systems are the most widely adopted \citep[as reviewed by][]{Bessell2005}.
\begin{itemize}
\item Vega magnitude system. The spectrum of Vega (or alternatively the average of several A0V stars, or the best fitting atmospheric model)
is used as the reference spectrum. The reference magnitudes are set so that Vega has zero magnitudes across all photometric bands. In some cases, a small offset could be applied, e.g., 0.03 for the Johnson system, or something else for other systems, following \citet{Cox2000} and \citet{Girardi2002}. By default, we use the latest Vega spectrum (\texttt{alpha\_lyr\_stis\_010.fits}) from the CALSPEC\footnote{\url{ftp://ftp.stsci.edu/cdbs/current_calspec/}} database \citep{Bohlin2014} here, while in \cite{Chen2019}, \href{ftp://ftp.stsci.edu/cdbs/current_calspec/}{\texttt{alpha\_lyr\_stis\_008.fits}} was used. We evaluated the impact of different reference Vega spectra used with the JWST wide filters (CSST/MSC and CSST/MCI are in AB magnitude system), finding a $\sim$ 0.01\,mag difference in the magnitudes and colours, which can be neglected for studies of high redshift stars with normally large photometric errors.

\item AB magnitude system \citep{Oke1974,Oke1983}. The reference spectrum has a constant flux of $f^0_\nu=10^\frac{48.60}{-2.5}$ $\rm erg\ s^{-1}\ cm^{-2}\ Hz^{-1}$. The reference magnitudes are therefore set to $-48.60$\,mag.
\end{itemize}

However, synthetic spectral libraries usually provide the stellar flux at the stellar surface instead of that at the receiver of a certain distance from the star. The stellar surface is defined at a given optical depth (e.g., $\tau=2/3$ or 20) for stellar atmosphere models assuming the plane-parallel geometry, or defined at a radius $R$ in the case of the spherical geometry. The stellar surface flux $F_\lambda$ is related to the effective temperature $T_{\rm eff}$ of the star by the Stefan-Boltzmann law 
\begin{equation}
  F_{\rm bol} \equiv \int_0^\infty F_\lambda \diff\lambda=\sigma T^4_{\rm eff},
  \label{eq_Fbol}
\end{equation}
where $\sigma$ is the Stefan-Boltzmann constant. By placing the star at 10\,pc from the telescope, the flux received is then
\begin{equation*}
  f_{\lambda, \rm 10\,pc}=\left(\frac{R}{\rm 10\,pc}\right)^2 F_\lambda 10^{-0.4A_\lambda},
\end{equation*}
where $A_\lambda$ is the assumed extinction between the star and the observer. This formula introduces the stellar radius $R$, which causes difficulty for synthetic spectral libraries in plane-parallel geometry with stellar radius $R$ undefined.

The absolute magnitude $M_i$ for a photon-counting photometric system is
\begin{equation}
  \begin{split}
    M_i&=-2.5{\rm log}\left[\frac{\int_{\lambda_1}^{\lambda_2} \lambda f_{\lambda, \rm 10\,pc} S_{\lambda, i} \diff\lambda}
      {\int_{\lambda_1}^{\lambda_2} \lambda f^0_\lambda S_{\lambda, i} \diff\lambda}\right] + m_{i,0}\\
    &=-2.5{\rm log}\left[ \left(\frac{R}{\rm 10\,pc}\right)^2
      \frac{\int_{\lambda_1}^{\lambda_2} \lambda F_\lambda 10^{-0.4A_\lambda} S_{\lambda, i} \diff\lambda}
           {\int_{\lambda_1}^{\lambda_2} \lambda f^0_\lambda S_{\lambda, i} \diff\lambda}\right] + m_{i,0}.
  \end{split}
  \label{eq_Mag}
\end{equation}
The definition of bolometric magnitude $M_{\rm bol}$ is
\begin{equation}
  \begin{split}
    M_{\rm bol} &\equiv M_{\rm bol,0} - 2.5{\rm log} (L/L_{\rm 0})\\
    &=M_{\rm bol,\odot} - 2.5{\rm log} (L/L_{\rm \odot})\\
    &= M_{\rm bol,\odot} - 2.5{\rm log} (4\pi R^2F_{\rm bol}/L_{\rm \odot})\\
    &= M_{\rm bol,\odot} - 2.5{\rm log} (4\pi R^2 \sigma\teff^4/L_{\rm \odot}).
  \end{split}
  \label{eq_Mbol}
\end{equation}
According to the IAU 2015 GAR B2 resolution \citep{IAU2015}, a radiation source with absolute bolometric magnitude $M_{\rm bol,0}=0$\,mag has a radiative luminosity of exactly $L_{\rm 0} = 3.0128 \times 10^{28}$ W. This is equivalent to stating that the absolute bolometric magnitude for the nominal solar luminosity ($L_{\rm \odot} = 3.828\times10^{26}$ W) is $M_{\rm bol,\odot}=4.74$\,mag.

Given an absolute magnitude $M_i$ or apparent magnitude $m_i$ in a given passband $i$ for a star of absolute bolometric magnitude $M_{\rm bol}$ or of apparent bolometric magnitude $m_{\rm bol}$, the bolometric correction $\BC_i$ is defined as:
\begin{equation}
  \BC_i\equiv M_{\rm bol}- M_i = m_{\rm bol}- m_i.
  \label{eq_BC}
\end{equation}

By combining equations (\ref{eq_Mag}), (\ref{eq_Mbol}) and (\ref{eq_BC}), we have 
\begin{equation} 
  \begin{split}
    \BC_i=&M_{\rm bol,\odot}-2.5{\rm log} \left( \frac{4\pi \sigma ({\rm 10\,pc})^2}{L_{\rm \odot}} \right) - 2.5{\rm log} (T^4_{\rm eff}) \\
    &+2.5{\rm log}\left[ \frac{\int_{\lambda_1}^{\lambda_2} \lambda F_\lambda 10^{-0.4A_\lambda} S_{\lambda, i} \diff\lambda}
      {\int_{\lambda_1}^{\lambda_2} \lambda f^0_\lambda S_{\lambda, i} \diff\lambda}\right] - m_{i,0}. 
  \end{split}
  \label{eq_BC_photon}
\end{equation}
The advantage of using the above equation to compute BCs is the elimination of the stellar radius $R$, which is undefined in plane-parallel atmosphere models. Therefore, once $F_\lambda$ (related to $T_{\rm eff}$ by equation (\ref{eq_Fbol})) for a star with given $T_{\rm eff}$, $\logg$ and metallicity ${\rm [M/H]}$ is provided, the corresponding $\BC_i$ can be derived.

Due to the progress in the development of the synthetic spectral libraries, nowadays $\BC_i$s are usually derived from synthetic spectral libraries and are provided in tabular, as functions of stellar parameters, such as $\teff$, $\logg$, metallicity $Z$, etc. By searching the BC tables according to a star's stellar parameters, we can obtain its bolometric correction $\BC_i(\teff, \logg, Z)$ in the passband $i$. Given the star's luminosity, its absolute magnitude in the passband $i$ can be derived by combining equations~(\ref{eq_Mbol}) and~(\ref{eq_BC}), yielding
\begin{equation}
  M_i = M_{\rm bol,\odot} - 2.5{\rm log} (L/L_{\rm \odot}) - \BC_i(\teff, \logg, Z).
  \label{eq_Magi}
\end{equation}
The application of this formula includes deriving the absolute magnitudes of stars from stellar evolutionary models.

When the distance $d$ to the star is known, the apparent magnitude can then be derived with
\begin{equation}
\begin{split}
  m_i &= M_i + 5{\rm log} \left(\frac{d}{\rm 10\,pc}\right) \\
      &= 5{\rm log}\left(\frac{d}{\rm 10\,pc}\right) + M_{\rm bol,\odot} - 2.5{\rm log} (L/L_{\rm \odot}) - \BC_i.
  \end{split}
  \label{eq_mag_BC}
\end{equation}
The application of this equation includes colour-magnitude diagram (CMD) fitting to the observations with theoretical models. Once the $\BC_i$ is known from theoretical models, the intrinsic luminosity can then be derived with the observed magnitude $m_i$ and distance $d$ from the equation~(\ref{eq_mag_BC}).

For AB magnitude systems with photon-counting devices, we can either convert $f^0_\nu$ to $f^0_\lambda$ and use equation (\ref{eq_BC_photon}), or use the following equation instead:
\begin{equation}
  \begin{split}
    \BC_i=&M_{\rm bol,\odot}-2.5{\rm log}\left( \frac{4\pi \sigma ({\rm 10\,pc})^2}{L_{\rm \odot}} \right) - 2.5{\rm log} (T^4_{\rm eff})\\
    &+2.5{\rm log}\left[ \frac{\int_{\nu_1}^{\nu_2} F_\nu 10^{-0.4A_\nu} S_{\nu, i} \diff\nu/\nu}
      {\int_{\nu_1}^{\nu_2} f^0_\nu S_{\nu, i} \diff\nu/\nu}\right] - m_{i,0}\\
      =&M_{\rm bol,\odot}-2.5{\rm log}\left( \frac{4\pi \sigma ({\rm 10\,pc})^2}{L_{\rm \odot}} \right) - 2.5{\rm log} (T^4_{\rm eff})\\
    &+2.5{\rm log}\left[ \frac{\int_{\nu_1}^{\nu_2} F_\nu 10^{-0.4A_\nu} S_{\nu, i} \diff\nu/\nu}
      {\int_{\nu_1}^{\nu_2} S_{\nu, i} \diff\nu/\nu}\right].
  \end{split}
  \label{eq_BC_photon_AB}
\end{equation}

\subsection{Non-zero redshift case}
In the case of cosmologically redshifted stars, the apparent magnitudes are affected by the cosmological dimming, the cosmological $K$-correction as well as the dust extinction. The cosmological dimming can be easily dealt with by replacing the distance $d$ in equation~(\ref{eq_mag}) with the cosmological distance $d_{\rm L}$, regardless of the passband considered and the spectral type of the star. The cosmological distance $d_{\rm L}$ is only fixed by the redshift and the adopted cosmology. Throughout the present work, we adopt the flat $\Lambda$CDM cosmology with parameters from \cite{Planck2018}, $h = H_0 {\rm [km\,s^{-1}\,Mpc^{-1} ]}/100 = 0.677$, $\Omega_\Lambda = 0.69$, $\Omega_{\rm M} = 0.31$. The cosmological $K$-correction comes from three effects: 1) the wavelength $\lambda$ of the photon received by the telescope is from a shorter wavelength of the rest-frame spectrum, which is $\lambda_{\rm e}=\lambda/(1+z)$; 2) the energy of the photon $h\nu$ received is emitted with a higher energy in the star's rest-frame, which is $h \nu_{\rm e}=h\nu*(1+z)$; and 3) the received number rate of the photon by the telescope is reduced by $(1+z)$ from that when the photons are emitted in the rest-frame. The combined effect of the cosmological dimming and cosmological $K$-correction results in the flux $f_{\lambda, z}$ observed by the telescope and the corresponding apparent magnitude $m_{i,z}$. We can write the formula for the apparent magnitude in the same form as equation~(\ref{eq_mag}) but with $f_{\lambda}$ replaced by $f_{\lambda,z}$,
\begin{equation}
  m_{i,z}=-2.5{\rm log}\left[\frac{\int_{\lambda_1}^{\lambda_2} \lambda f_{\lambda,z} S_{\lambda, i} \diff\lambda}
    {\int_{\lambda_1}^{\lambda_2} \lambda f^0_\lambda S_{\lambda, i} \diff\lambda}\right] + m_{i,0}.
    \label{eq_mag_z}
\end{equation}
Notice that $f^0_\lambda$ and $m_{i,0}$ remain unchanged. Suppose the observed flux $f_{\lambda, z}$ comes from a star with flux $F_{\lambda_{\rm e}}$ at its rest-frame wavelength $\lambda_{\rm e}$ emitted at its surface of radius $R$, this rest-frame flux is related to its effective temperature $\teff$ by
\begin{equation}
  F_{\rm bol} \equiv \int_0^\infty F_{\lambda_{\rm e}} \diff\lambda_{\rm e} = \sigma T^4_{\rm eff}.
  \label{eq_Fbol_z}
\end{equation}
This is the same as equation~(\ref{eq_Fbol}), since these quantities are the star's intrinsic properties and are in the star's rest-frame. However, the relation between the observed flux $f_{\lambda, z}$ and $F_{\lambda_{\rm e}}$ is changed to
\begin{equation}
  \begin{split}
  f_{\lambda, z}\diff\lambda &=\left(\frac{R}{d_{\rm L}}\right)^2 F_{\lambda_{\rm e}} 10^{-0.4\left(\alpha_{\lambda_{\rm e}}+A_\lambda\right)} \diff\lambda_{\rm e} \\
  &=\left(\frac{R}{d_{\rm L}}\right)^2 F_{\lambda/(1+z)} 10^{-0.4\left(\alpha_{\lambda/(1+z)}+A_\lambda\right)} \frac{\diff\lambda}{1+z},
  \end{split}
  \label{eq_flam_z}
\end{equation}
where $\alpha_{\lambda_{\rm e}}$ and $A_\lambda$ denote the extinction due to the star's local dust (denoted as ``environmental dust'' here, including the circumstellar dust and the dust of the host galaxy) and the Galactic dust, respectively. The extinction curves for these components can be quite different. Notice that the extinction due to the intervening galactic dust and IGM along the star's light travel is not considered here for simplicity. 

Without ambiguity, we define the ``absolute flux'' corresponding to the observed flux $f_{\lambda, z}$ as 
\begin{equation}
  f_{\lambda, z,10\,{\rm pc}}\equiv f_{\lambda, z}\left(\frac{d_{\rm L}}{10\,{\rm pc}}\right)^2 
  = \left(\frac{R}{\rm 10 pc}\right)^2 10^{-0.4\left(\alpha_{\lambda/(1+z)}+A_\lambda\right)} \frac{F_{\lambda/(1+z)}}{1+z}.
  \label{eq_flam_z10}
\end{equation}
This definition puts a star at 10\,pc from the receiver, with the same spectral shape as that of the redshifted one but keeping the same bolometric luminosity. Inserting this flux into equation~(\ref{eq_mag_z}), we have the formula for the ``absolute magnitude''
\begin{equation}
  \begin{split}
    M_{i,z} =& -2.5{\rm log}\left[\frac{\int_{\lambda_1}^{\lambda_2} \lambda f_{\lambda, z, {\rm 10\,pc}} S_{\lambda, i} \diff\lambda}
      {\int_{\lambda_1}^{\lambda_2} \lambda f^0_\lambda S_{\lambda, i} \diff\lambda}\right] + m_{i,0}\\
    =& -2.5{\rm log}\left[ \left(\frac{R}{\rm 10\,pc}\right)^2
      \frac{\int_{\lambda_1}^{\lambda_2} \lambda \frac{F_{\lambda/(1+z)}}{1+z}10^{-0.4\left(\alpha_{\lambda/(1+z)}+A_\lambda\right)} S_{\lambda, i} \diff\lambda}
           {\int_{\lambda_1}^{\lambda_2} \lambda f^0_\lambda S_{\lambda, i} \diff\lambda}\right] \\
           &+ m_{i,0}.
  \end{split}
  \label{eq_Mag_z}
\end{equation}

We have the same definition of bolometric magnitude for $M_{\rm bol,z}$ as eq. (\ref{eq_Mbol}),
\begin{equation}
    M_{\rm bol,z} \equiv M_{\rm bol,z=0} = M_{\rm bol,\odot} - 2.5{\rm log} (4\pi R^2 \sigma\teff^4/L_{\rm \odot}).
  \label{eq_Mbol_z}
\end{equation}
This is equivalent to stating that the bolometric magnitude for a cosmologically redshifted star is defined in its rest-frame.

The bolometric correction $\BC_i$ is defined similarly to eq. (\ref{eq_BC})
\begin{equation}
  \BC_{i,z}\equiv M_{\rm bol,z}- M_{i,z}.
  \label{eq_BC_z}
\end{equation}

By combining equations (\ref{eq_Mag_z}), (\ref{eq_Mbol_z}), and (\ref{eq_BC_z}), we have 
\begin{equation} 
  \begin{split}
    \BC_{i,z}=&M_{\rm bol,\odot}-2.5{\rm log} \left( \frac{4\pi \sigma ({\rm 10\,pc})^2}{L_{\rm \odot}} \right) - 2.5{\rm log} (T^4_{\rm eff}) \\
    &+2.5{\rm log}\left[ \frac{\int_{\lambda_1}^{\lambda_2} \lambda \frac{F_{\lambda/(1+z)}}{1+z} 10^{-0.4\left(\alpha_{\lambda/(1+z)}+A_\lambda\right)} S_{\lambda, i} \diff\lambda}
      {\int_{\lambda_1}^{\lambda_2} \lambda f^0_\lambda S_{\lambda, i} \diff\lambda}\right] - m_{i,0}. 
  \end{split}
  \label{eq_BC_photon_z}
\end{equation}
Compared to equation~(\ref{eq_BC_photon}), the differences are the $F_{\lambda/(1+z)}$, $\frac{1}{1+z}$ and extinction terms in the integrand. The $F_{\lambda/(1+z)}$ term specifies that we have to use the rest-frame flux at $\lambda/(1+z)$ for the passband wavelength point $\lambda$, while the $\frac{1}{1+z}$ term specifies the stretch of the waveband due to the redshift.

The difference between $\BC_{i}$ and $\BC_{i,z}$ gives the cosmological $K$-correction, which is a familiar concept in the research field of high-redshift galaxies. It is expressed as
\begin{equation} 
  \begin{split}
    K_{i,z}=&\BC_{i,z}-\BC_{i,z=0}\\
    =&2.5{\rm log}\left[ \frac{\int_{\lambda_1}^{\lambda_2} \lambda \frac{F_{\lambda/(1+z)}}{1+z} 10^{-0.4\left(\alpha_{\lambda/(1+z)}+A_\lambda\right)} S_{\lambda, i} \diff\lambda}
      {\int_{\lambda_1}^{\lambda_2} \lambda F_\lambda 10^{-0.4\left(\alpha_{\lambda/(1+z)}+A_\lambda\right)} S_{\lambda, i} \diff\lambda}\right].
  \end{split}
  \label{eq_K}
\end{equation}

The counterpart of eq.~(\ref{eq_BC_photon_z})for the AB magnitude system is 
\begin{equation}
  \begin{split}
    \BC_{i,z}=&M_{\rm bol,\odot}-2.5{\rm log}\left( \frac{4\pi \sigma ({\rm 10\,pc})^2}{L_{\rm \odot}} \right) - 2.5{\rm log} (T^4_{\rm eff})\\
    &+2.5{\rm log}\left[ \frac{\int_{\nu_1}^{\nu_2} (1+z)F_{\nu(1+z)} 10^{-0.4\left(\alpha_{\nu(1+z)}+A_\nu\right)} S_{\nu, i}  \diff\nu/\nu}
      {\int_{\nu_1}^{\nu_2} f^0_\nu S_{\nu, i} \diff\nu/\nu}\right] - m_{i,0}. 
  \end{split}
  \label{eq_BC_photon_AB_z}
\end{equation}

Once we have computed $\BC_{i,z}$, we can compute the ``absolute magnitude'' as in equation~(\ref{eq_Magi}) and compute the apparent magnitude as in equation~(\ref{eq_mag_BC}) but with $d$ replaced by $d_{\rm L}$. 

For high redshift studies, it is sometimes useful to use the flux instead of the magnitude. This flux can be easily derived from the magnitude with the reference spectrum as 
\begin{equation}
    \mathcal{F}_{i,z}
    =10^{(m_{i,0}-m_{i,z})/2.5}\mathcal{F}_{{\rm Ref},i}.
  \label{eq_flux_z}
\end{equation}

The reference flux for the Vega magnitude system is computed as
\begin{equation} 
   \mathcal{F}_{{\rm Ref},i}= \mathcal{F}_{{\rm Vega},i}=\frac{\int_{\lambda_1}^{\lambda_2} \lambda f^0_\lambda S_{\lambda, i} \diff\lambda}
    {\int_{\lambda_1}^{\lambda_2} \lambda S_{\lambda, i} \diff\lambda}.
  \label{eq_flux_vega_photon_z}
\end{equation}
This flux changes with different passband transmission curves and Vega reference spectra.

For the AB magnitude system, the flux for the reference spectra is a constant as 
\begin{equation} 
   \mathcal{F}_{{\rm Ref},i}= \mathcal{F}_{{\rm AB},i} 
   =10^\frac{48.60}{-2.5} {\rm erg\ s^{-1}\ cm^{-2}\ Hz^{-1}}.
  \label{eq_flux_ab_photon_z}
\end{equation}

\section{Synthetic stellar spectral libraries for cosmologically redshifted stars}
\label{sec:spec}

In this section, we introduce the synthetic stellar spectral library used for computing the BCs for cosmologically redshifted stars.

\subsection{Stellar spectral libraries}
Besides the stellar spectral libraries presented in \cite{Chen2019}, we also include the libraries computed with non-local thermodynamic equilibrium (non-LTE) models for hot stars. In the following, we summarise the main properties of these stellar spectral libraries, which mostly concern massive stars.

    CK03\footnote{\url{http://wwwuser.oats.inaf.it/castelli/grids.html}} is a set of models widely used in the community computed with the plane-parallel atmosphere code \texttt{ATLAS9} and the spectrum synthesis code \texttt{SYNTHE} \citep{ATLAS9}. Though being a set of LTE models without taking into account the effect of stellar wind, it covers a wide range of $\teff$ (3500\,K to 50000\,K), $\logg$ (0.0 dex to 5.0 dex, in cgs units) and metallicity ($\mh$ from -5.5 to +0.5) and can be used for comparison with non-LTE models and those with stellar wind. These models are based on the solar abundances from \cite{GS98} and make use of set of molecular absorption lines including TiO and H$_2$O, as well as absorption lines from quasi-molecular H-H and H-H$^+$. In addition to the assumption of LTE, another drawback of this model set is its low spectral resolution with only 1221 grid points that cover the wavelength range from 90.9\,\AA\, to 160 $\mu$m. This limits the usage of the models to only broad band photometry.

    PHOENIX/LYON/BT-Settl\footnote{\url{https://archive.stsci.edu/hlsps/reference-atlases/cdbs/grid/phoenix/}} is another set of widely used stellar spectral library and is computed with the \texttt{PHOENIX} code \citep{Allard1995,Allard1997,PHOENIX}. Among the various suites of the \texttt{PHOENIX} models, we use the BT-Settl LTE models for its more complete coverage of stellar parameters. The BT-Settl models use the \citet{AGSS2009} solar abundances and are provided for $2600\,$K $\leq \teff < 50000\,K$, $0.5 < \logg < 6$, and metallicities $-4.0\lesssim {\rm [M/H]} \lesssim +0.5$. The spectra are computed with very high resolution and cover the wavelength range from 10\,\AA\, to 1\,mm.

    PoWR\footnote{\url{https://www.astro.physik.uni-potsdam.de/~wrh/PoWR/powrgrid1.php}} Wolf-Rayet (WR) models is one of the most extensive sets of models for WR stars \citep{PoWR2002, PoWR2004, PoWR2012, PoWR2015}, computed with the \texttt{PoWR} code. These models adopt the solar abundances from \citet{GS98} and are computed for the metallicities of the Milky Way (MW), Large Magellanic Cloud (LMC), Small Magellanic Cloud (SMC), and sub-SMC. The WR models include late-type WN models of different hydrogen fractions, early-type WN models, and WC models. WR stars typically have wind densities one order of magnitude larger than those of massive O-type stars, and therefore the transformed radius $\log R_\mathrm{t}$ is introduced as a better parameter than $\logg$ \citep{Schmutz1989}. The transformed radius is defined as
    \begin{equation} 
    R_\mathrm{t}=R_* \left(\frac{v_\infty}{\rm 2500~ km\,\mathrm{s}^{-1}} \middle/ 
    \frac{\dot{M}\sqrt{D}}{10^{-4}~{\rm M_\odot\,\mathrm{yr}^{-1}}}\right)^{2/3}, 
    \end{equation} 
    with $R_*$, $v_\infty$, $\dot{M}$ and $D$ being the radius where the Rosseland continuum optical depth is 20, terminal velocity, mass-loss rate and clumping factor, respectively. This calls for attention when interpolating stellar evolutionary models with that of PoWR WR atmosphere models. The stellar evolutionary models usually do not provide the terminal velocity and clumping factor. Therefore, it is better to eliminate them in the above definition for the purpose of interpolating stellar evolutionary models with the PoWR library. Since each model grid of PoWR model set is characterised by a set of common parameter settings, such as stellar luminosity, terminal wind velocity, clumping constant, and chemical composition, it does not change the situation that models of the same $R_\mathrm{t}$ have similar spectra. Therefore, instead of $R_\mathrm{t}$ we define $R_\mathrm{n}$ as 
    \begin{equation} 
    \begin{split}
    R_\mathrm{n} &\equiv R_\mathrm{t}\left( \frac{v_\infty}{\rm 2500~ km\,\mathrm{s}^{-1} \sqrt{D}} \right)^{-2/3} \\
    &=R_* \left(\frac{\dot{M} }{10^{-4}~{\rm M_\odot\,\mathrm{yr}^{-1}}}\right)^{-2/3}, 
    \end{split}
    \end{equation} 
    At $R_*$, the corresponding effective temperature $T_*$ is defined.
    $T_*$ and $R_*$ are connected with the model luminosity $L_*$ using the Stefan-Boltzmann law.
    Notice that here $T_*$ is defined at the layer where the Rosseland continuum optical depth is 20 rather than the total Rosseland optical depth, which should be larger than the continuum one due to the contribution of line opacities though the line contribution could be small due to the high temperature. Also notice that $T_*$ defined in PoWR is different from that of the normal $\teff$ which is defined at the radius where the Rosseland optical depth is 2/3. The PoWR WR model sets are regularly spaced in the $T_*$ and $\log R_\mathrm{t}$. The WR star models are computed for $\logteff$ from $\sim$ 4.4 to $\sim$ 5.3 and $\log R_\mathrm{t}$ from $-0.8$ to $1.9$. The high resolution spectra cover the wavelength range from 200\,\AA\, to 8\,$\mu$m.

    PoWR O, B star model set represents one of the most extensive O, B star models with stellar wind \citep{PoWR2019}. Similar to the PoWR WR models described above, the O, B star models are also computed for the metallicities of the MW, LMC, SMC and sub-SMC and for $T_*$ (effective temperature where the Rosseland continuum optical depth is 20) from 15,000\,K to $\sim$ 50,000\,K and for $\logg$ (unlike the $\log R_\mathrm{t}$ used for PoWR WR models) from 2.0 to 4.4. For each $T_*$ and $\logg$ of each metallicity, models of different mass-loss rates are provided. For all metallicities, models with 1/3 of the \cite{Vink2001} mass-loss recipe are computed. According to the eqs. (24) and (25) of \cite{Vink2001}, the mass-loss rate is a function of the metallicity, stellar luminosity, stellar mass, effective temperature, and terminal velocity. For MW and LMC metallicity, models of fixed $\dot{M}=10^{-7} \Msun/{\rm yr}$ are also computed. For SMC metallicity, models with three fixed wind strength parameter ${\rm log} Q$ values (-13, -12, and -14) are computed, where
    \begin{equation}
    Q=\left(\frac{\dot{M}}{\Msun/{\rm yr}}\sqrt{D}\right)/\left(\frac{v_\infty}{\rm km/s} \frac{R_*}{\Rsun}\right).
    \end{equation}
    With fixed ${\rm log} Q$ values, the mass-loss rate parameters are non-constant throughout the grids. We download both the low spectral resolution spectral energy distribution (SED) and the high spectral resolution synthetic spectra from the PoWR web interface and combine them to generate the PoWR spectral library covering a broad enough wavelength range and with high spectral resolution. Especially, the EUV band (200 \AA\, to 950 \AA) spectra are newly computed for this work.
    
    TLUSTY\footnote{\url{http://tlusty.oca.eu}} models are the non-LTE, plane-parallel, hydrostatic model atmospheres for O, B stars, computed with the stellar atmosphere model code \texttt{TLUSTY} and the spectral synthesis code \texttt{Synspec}. These models are computed with the solar abundance from \cite{GS98}. The set of O star models is computed at ten metallicities from metal-free to twice the solar metallicity, $\teff$ from 27,500\,K to 55,000\,K with a 2,500\,K step, $\logg$ from 3.0 to 4.75 with a 0.25 dex step, and the microturbulent velocity of 2 ${\rm km s^{-1}}$ \citep{Lanz2003}. The set of B star models is computed at six metallicities, from metal-free to twice the solar metallicity, $\teff$ from 15,000\,K to 30,000\,K with a 1,000\,K step, $\logg$ from 1.75 to 4.75 with a 0.25 dex step, and the microturbulent velocity (2 ${\rm km s^{-1}}$) \citep{Lanz2007}. In addition, the model atmospheres for B-type supergiants ($\logg \leq 3.0$) have been calculated with a higher microturbulent velocity of 10 ${\rm km s^{-1}}$ and a surface composition that is enriched in helium and nitrogen and depleted in carbon \citep{Lanz2007}. The spectra cover the wavelength range from 54\,\AA\,to 300\,$\mu$m.

    CMFGEN\footnote{\url{https://sites.pitt.edu/~hillier/web/CMFGEN.htm}} models. The \texttt{CMFGEN} code is one of the most sophisticated codes for computing the non-LTE, line-blanketed, expanding atmosphere models with stellar wind and for computing emergent spectra for hot stars \citep{Hillier1998,Hillier2012}. \cite{Marcolino2024} presented the largest \texttt{CMFGEN} models for O, B stars, with 272 models for LMC metallicity and 305 models for SMC metallicity. These models used the solar abundance from \cite{AGSS2009} and are computed at 1/2 LMC and 1/5 SMC solar metallicity for LMC and SMC, respectively. We download the spectra for these models from the Pollux database \citep{Palacios2010}. The spectra cover the wavelength range from 50\,\AA\,to 25\,$\mu$m. Furthermore, \cite{Martins2017,Martins2021,Martins2022} presented \texttt{CMFGEN} spectra for 245 main-sequence star models at several metallicities. However, at each metallicity, the number of models is less than 100 and the models are irregularly distributed in $\teff$ and $\logg$ space. So we discard these models for the moment. The Pollux database also presents 11 CMFGEN models for WR stars, which are not considered here. 

    WM-basic \citep{Chen2015} model set includes models computed with the \texttt{WM-basic}\footnote{\url{https://www.usm.uni-muenchen.de/people/adi/Programs/Programs.html}} interface to the program package to model the atmosphere of hot stars with stellar wind \citep{Pauldrach1986}. The modelling takes into account both the effects of extended winds and those of non-LTE, which significantly affect the emergent spectra of hot stars. The current version of the spectral library of \cite{Chen2015} contains models for metallicity $Z=$[0.02, 0.008, 0.004, 0.001, 0.0005, 0.0001] with the same solar chemical composition pattern as that of PARSEC models. At each metallicity, the models are with $\logte$ = 4.3 to 5.0 in a 0.025\,dex step. At each $\teff$, the values of $\logg$ are considered in 0.5\,dex step, with the upper $\logg$ boundary corresponding to the Eddington limit and the lower boundary determined by the fact that the line-driven radiation force cannot initiate the stellar wind. At each $\teff$ and $\logg$ grid point, five different mass-loss rates are considered, namely $\dot{M}=[10^{-7},10^{-6},10^{-5},10^{-4},10^{-3}]\,{\rm M}_{\odot}\,\mathrm{yr}^{-1}$. Such a regular gridding enables efficient interpolation with stellar evolutionary models. This library contains about 3000 models. These model sets have already been included in works for providing spectra for stellar population synthesis models \cite{Lecroq2024}. The spectra cover the wavelength range from 50\,\AA\,to 60\,mm.

Other libraries, such as the COMARCS one, are not presented here, since they are not related to massive stars but to asymptotic giant branch stars (AGBs). The PHOENIX/G\"{o}ttingen models are not included as they provide models only up to $\teff$ 12,000\,K. In table~\ref{table:1}, we summarise the coverage of the main stellar parameters of the above spectral libraries.

\begin{table*}[ht!]
    \caption{Properties of the spectral libraries.}
    \label{table:1}
    \centering
    \begin{tabular}{c|c|c|c|c}\hline
         Library name & Metallicity & Teff range & logg or logRt range & wavelength range\\ \hline 
         CK03         & -5.5 to 0.5 & 3500\,K to 50\,kK & 0 to 5& 90 \AA\, to 160 $\mu$m\\
         PHOENIX/LYON/BT-Settl & -4 to 0.5 & 2600\,K to 50\,kK & 0.5 to 6 & 10 \AA\, to 1\,mm\\
         PoWR WR & -- & 4.4 to 5.3 & -0.8 to 1.9 & 200 \AA\, to 8\,$\mu$m\\
         PoWR OB & 0.013 to 0.0004& 15\,kK to 50\,kK & 2 to 4.4 & 4.9 \AA\, to 84\,mm\\
         TLUSTY & 0 to 0.034 & 15\,kK to 55\,kK& 1.75 to 4.75 & 54 \AA\, to 300\,$\mu$m\\
         CMFGEN & -0.3,-0.73 & $\sim$23\,kK to 55\,kK & $\sim$2.5 to 4.4 & 50 \AA\, to 25\,$\mu$m\\
         WM-basic & 0.0001 to 0.02 & 4.3 to 5 & 2.5 to 6 & 50 \AA\, to 60\,mm\\ \hline
    \end{tabular}
    \tablefoot{Summary of the properties of the spectral libraries used in this work. The values in the metallicity column are either in [M/H] (where a minus sign is given) or Z.}
\end{table*}

\begin{figure}[ht!]
    \centering
    \includegraphics[trim=2.2cm 0.8cm 2.2cm 2cm, clip, width=1\linewidth]{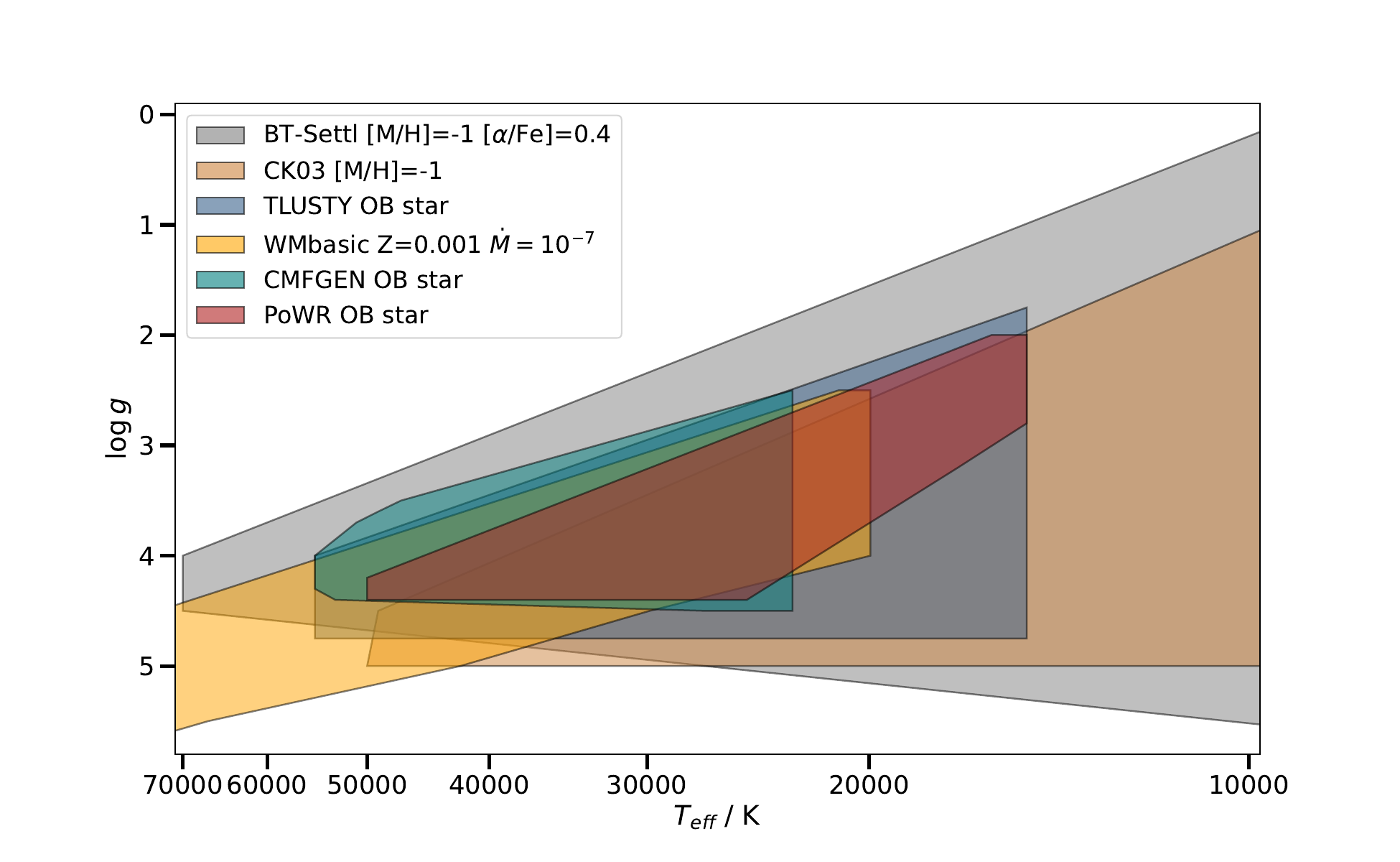}
    \caption{Parameter coverage of example atmosphere models on the Kiel diagram.}
    \label{fig:HR-lib}
\end{figure}

In figure~\ref{fig:HR-lib}, we show the distribution of the models (with $\teff > 10\,000$\,K) in the $\teff$-$\logg$ plane ($\teff$-${\rm log}R_\mathrm{n}$ for PoWR WR models). Although the PHOENIX/LYON/BT-Settl model set covers the widest range on this plane, the models are LTE or without mass-loss. Overall, these stellar spectral libraries provide decent coverage for stellar evolutionary models.

\subsection{Redshifted stellar spectra}
With the aforementioned stellar spectral libraries, we can compute the stellar spectra for stars at any given redshift. 

\begin{figure}
    \centering
    \includegraphics[trim=0cm 0cm 0cm 0cm, clip, width=0.99\linewidth]{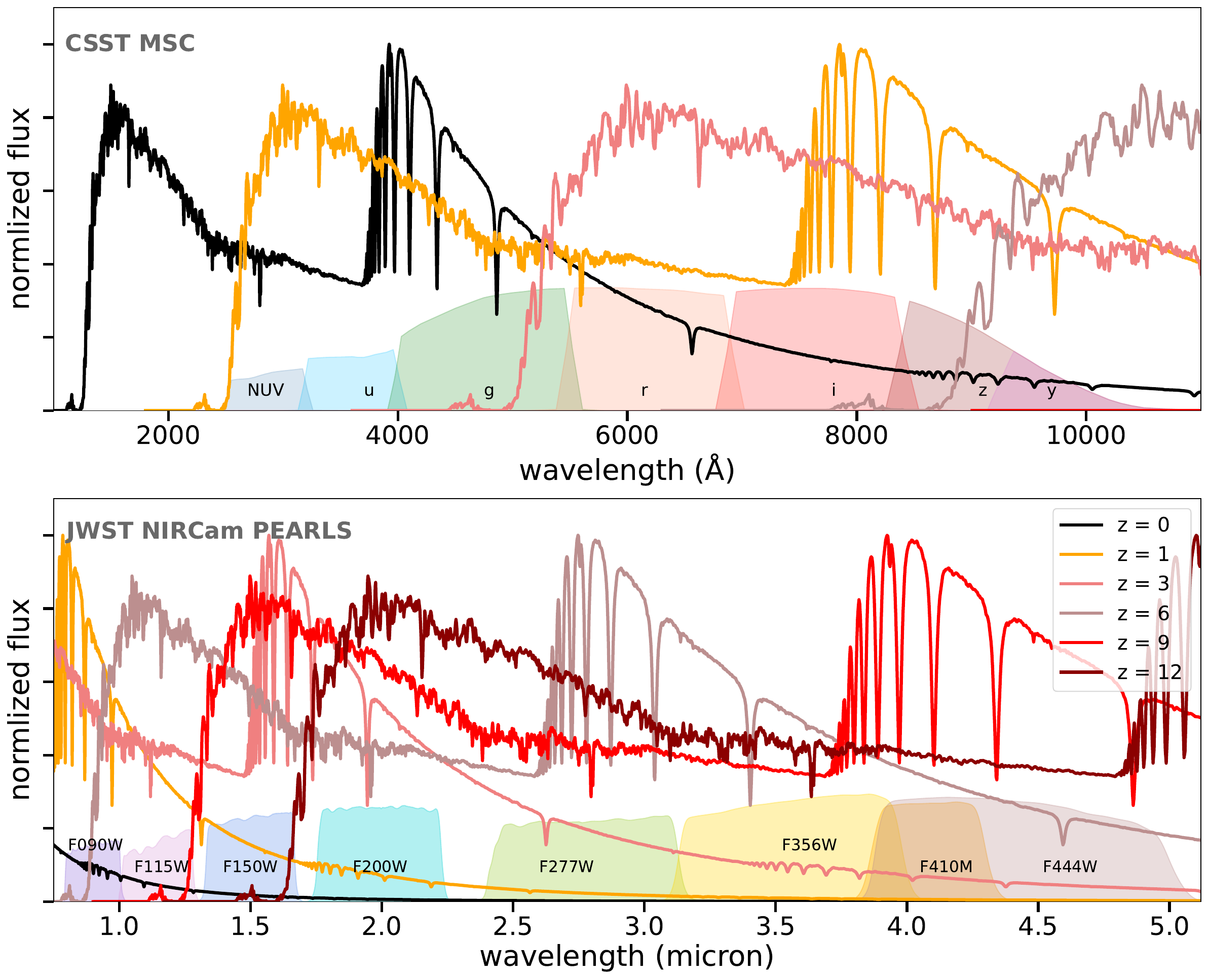}
    \caption{Normalised Vega spectrum (\texttt{alpha\_lyr\_stis\_010.fits}) at different redshifts. Transmission curves of the CSST/MSC filters (the upper panel: NUV, u, g, r, i, z, y), together with the JWST NIRCam filters used in the PEARLS \citep{pearls2023} project (the lower panel: F090W, F115W, F150W, F200W, F277W, F356W, F410M, F444W) are also illustrated. The spectra plotted are normalised, so the cosmological dimming effect brought by the huge luminosity distance is not displayed. }
    \label{fig:z_csst_jwst}
\end{figure}

In figure~\ref{fig:z_csst_jwst}, we display the Vega spectrum at various redshifts, with the CSST/MSC and JWST/NIRCam transmission curves overplotted. The plot shows the progressive redshift of the overall spectra shape and the prominent spectral features, such as the Balmer break. These features enable the redshift of cosmologically redshifted stars to be determined with photometric redshift methods as those for high redshift galaxies. With the CSST/MSC, the Balmer break can be well sampled across the redshift range from z=0 to $\sim$1, whereas the JWST/NIRCam enables sampling from z$\sim$2 up to $\gtrsim$9. Synergy between the CSST and JWST observations will provide a comprehensive picture of high-redshift stellar populations.

\begin{figure}
    \centering
    \includegraphics[trim=0cm 0cm 0cm 0cm, clip, width=0.99\linewidth]{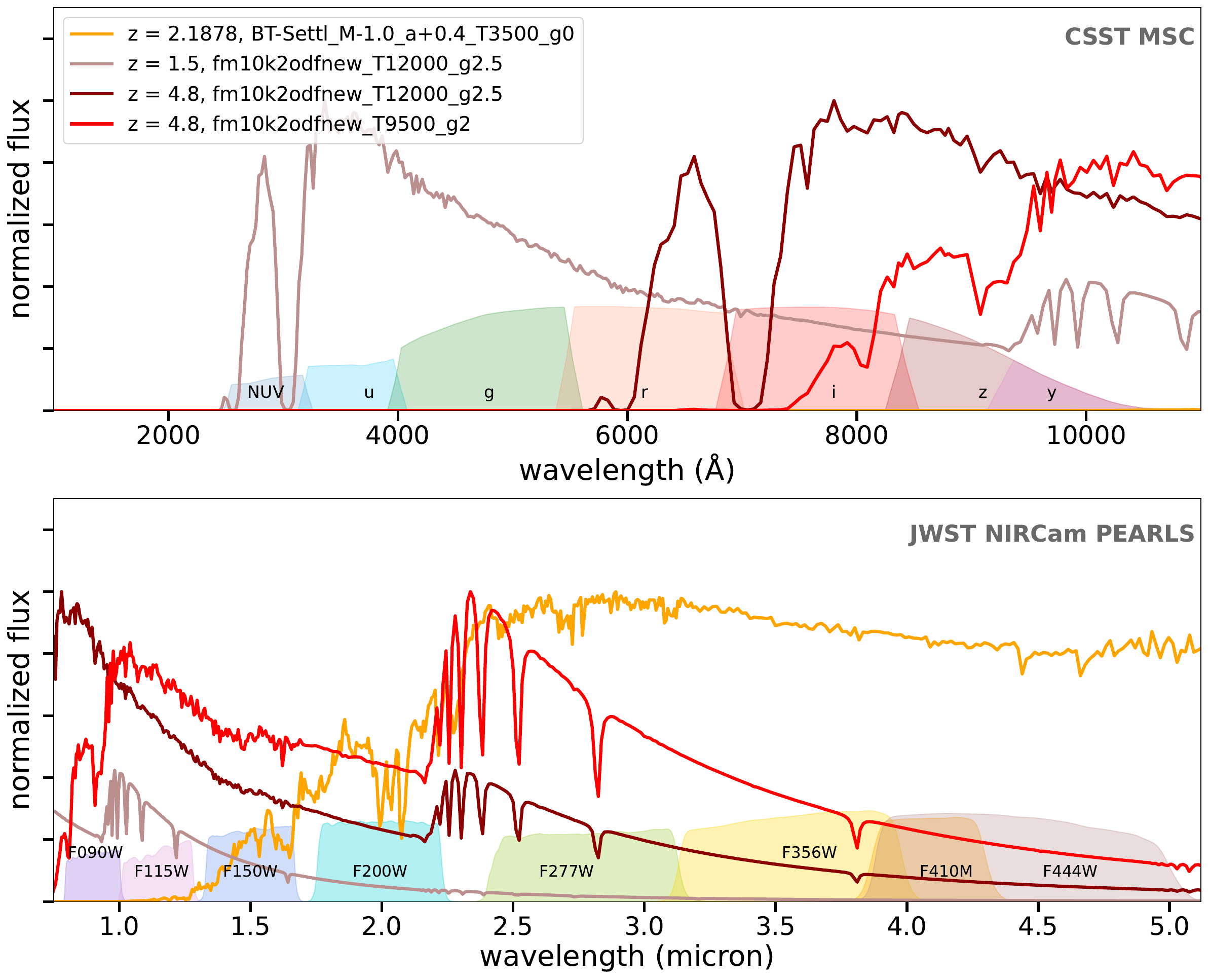}
    \caption{Normalised model spectra at different redshifts. The yellow spectrum with ``z=2.1878" is from BT-Settl library with [M/H]=-1.0, [$\alpha$/Fe]=0.4, $\teff = 3,500\,$K and $\logg=0$, corresponding to the $z\sim2.1878$ red supergiant star candidate by \cite{Diego2023}. The light brown spectrum with ``z=1.5" is from CK03 with [M/H]=-1.0, $\teff = 12,000\,$K and $\logg=2.5$, corresponding to the $z\sim1.5$ stellar candidate as first discovered by \cite{Kelly2018}. The brown spectrum with ``z=4.8" is from CK03 with [M/H]=-1.0, $\teff = 12,000\,$K and $\logg=2.5$, corresponding to the $z\sim4.8$ stellar candidate ``star-2'' by \cite{Meena2023} ([M/H]=-1). The red spectrum with ``z=4.8" is from CK03 with [M/H]=-1.0, $\teff = 9,500\,$K and $\logg=2.0$, corresponding to the $z\sim4.8$ stellar candidate ``star-1'' by \cite{Meena2023} ([M/H]=-1). The BT-Settl spectrum is smoothed for better illustration.}
    \label{fig:z_csst_jwst_model}
\end{figure}

In figure~\ref{fig:z_csst_jwst_model}, we show several model spectra with different $\teff$ and $\logg$ values at various redshifts. The parameters for each spectrum correspond to high redshift star candidates selected from the literature.

\subsection{Extinction curve}
To compute the BCs with equations (\ref{eq_BC_photon_z}) and (\ref{eq_BC_photon_AB_z}), extinction curves are needed. We use the \texttt{extinction}\footnote{\url{https://extinction.readthedocs.io}} Python routine, which includes most of the popular extinction laws in the literature. For the Galactic extinction, we take \citet{Cardelli1989} plus the \citet{O'Donnell1994} extinction law (CCM+O94) with $R_V=3.1$, to maintain consistency with the PARSEC CMD web page. For environmental extinction of the cosmologically redshifted stars, we adopt the one from \citet{Calzetti2000}, which is developed for galaxies with spectra dominated by massive stars. BCs with other extinction laws or with other $R_V$ values can be computed upon request. As discussed in \cite{Chen2019}, the attenuation in a given band is a non-linear function of the total extinction. For the broad bands, the convolution of the extinction with the spectra is necessary for more accurate extinction modelling, especially in the case of spectra with strong lines or sharp spectral features. We provide tables with Galactic extinction of 0, 0.1, 0.2, 0.3, 0.4, and 0.5. Since cosmologically redshifted star observations are usually carried out in low Galactic extinction areas, these values suffice for the application. For the environmental extinction of the cosmologically redshifted stars, we compute tables with values of 0, 0.5, 1.0, 2.0, 5.0. Tables with other values can be provided upon request. The extinction laws as reviewed by \citet{Salim2020} can be easily implemented.

\section{BCs with redshift}
\label{sec:bc}

Based on the aforementioned stellar spectral libraries, we employ a modified version of the \texttt{YBC} code \citep{Chen2019} to calculate the BCs at various redshifts. The YBC database provided by the original \texttt{YBC} code includes BCs for most of the widely used photometric systems. However, in this study, we only provide BCs for a limited set of photometric systems. This is primarily because, on the one hand, the HST, JWST, and CSST are expected to allocate significant observational resources to study cosmologically redshifted stars. On the other hand, the inclusion of additional parameters, namely redshift and environmental extinction, greatly increases the computational time and storage requirements. As a result of these constraints, we compute the BCs only for a few sets of filters and only at fixed values of redshift and environmental extinction. We refer to this database as the zYBC database and the tables can be downloaded via our website. 

\begin{figure}
    \centering
    \includegraphics[width=\linewidth]{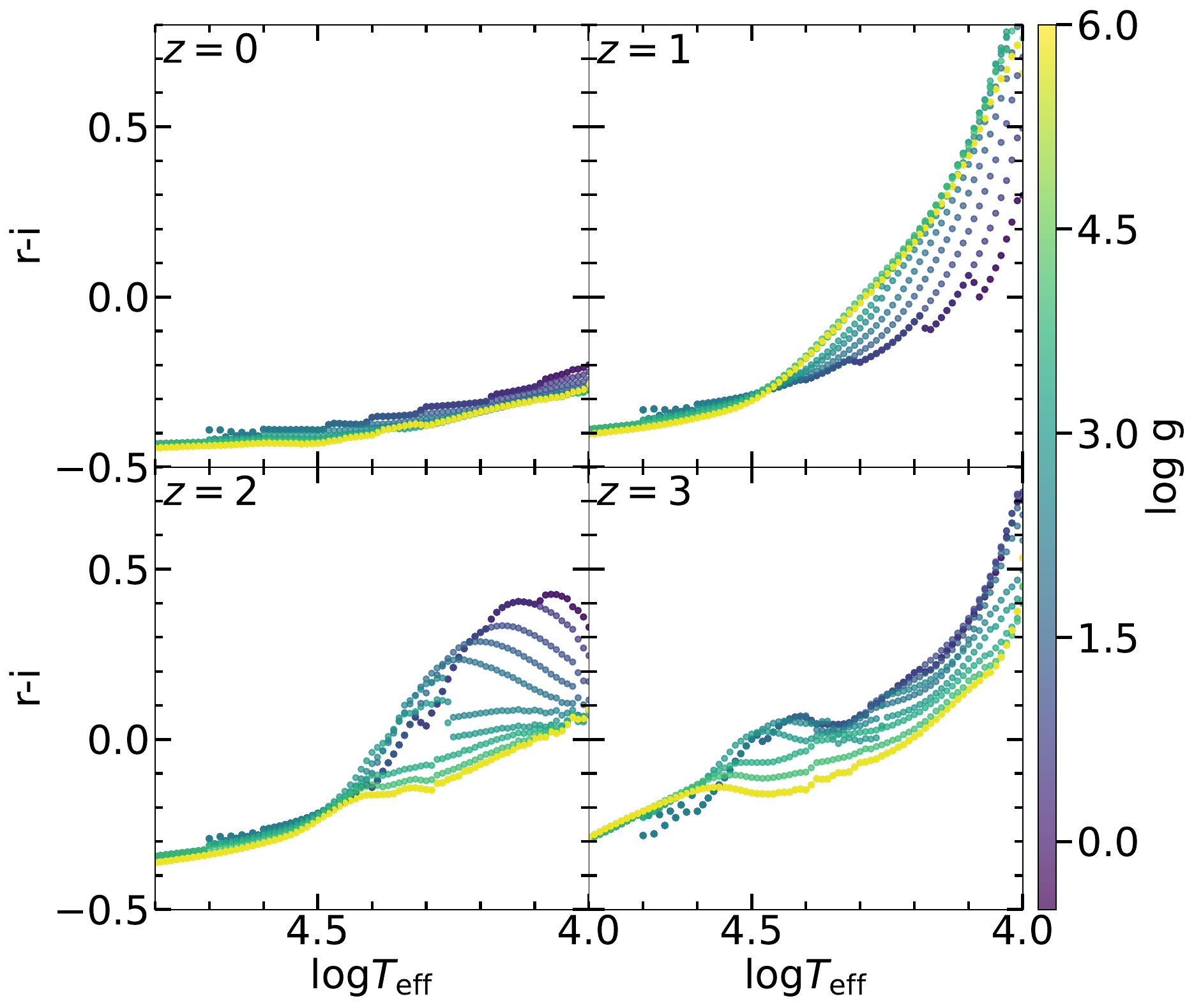}
    \caption{$r-i$ as a function of $\teff$ for the CSST/MSC for the Phoenix library.}
    \label{fig:color_teff_phoenix_csst_ri}
\end{figure}

\begin{figure}
    \centering
    \includegraphics[width=\linewidth]{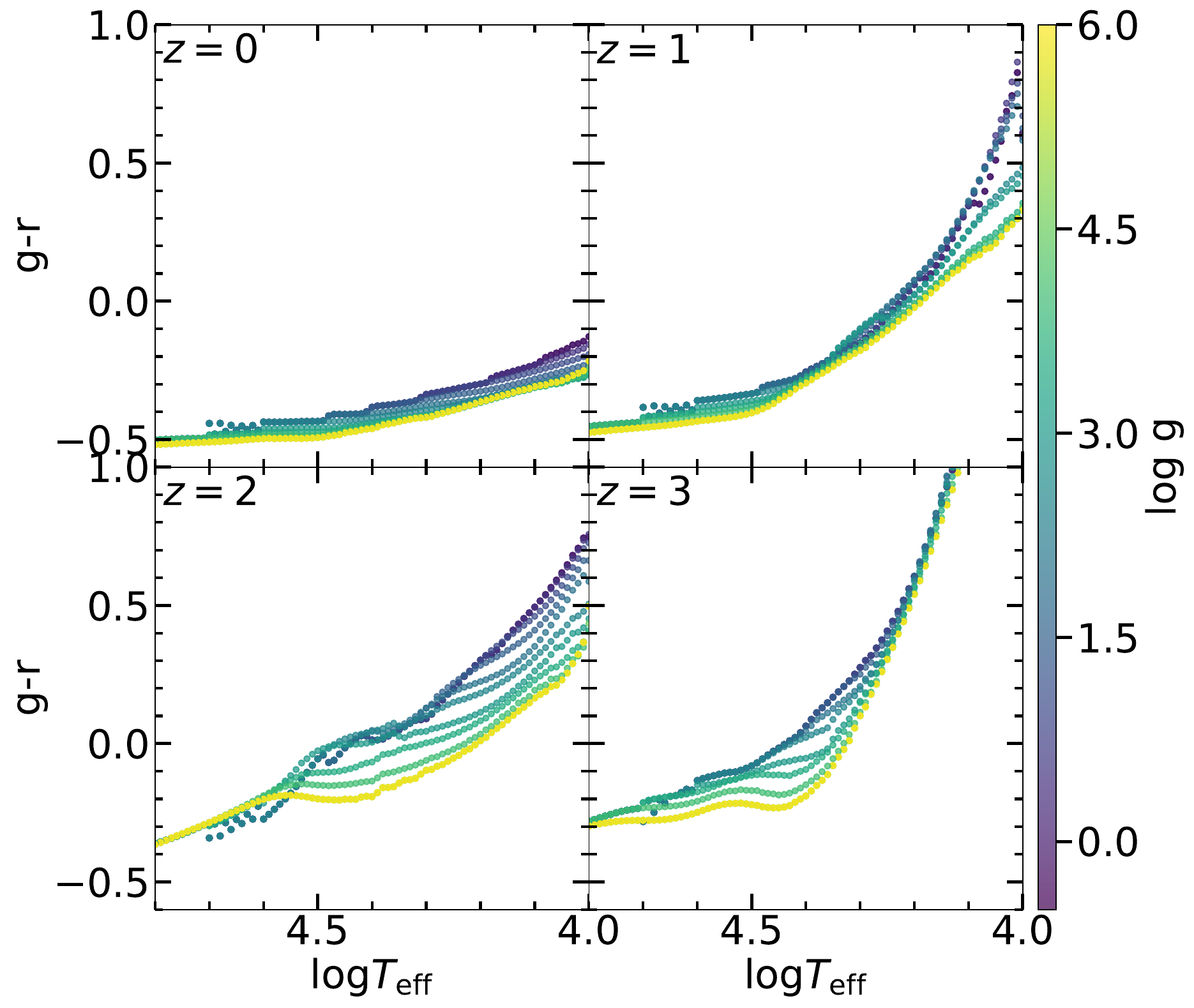}
    \caption{CSST/MSC $g-r$ as a function of $\teff$ for the Phoenix/BT-Settl library.}
    \label{fig:color_teff_phoenix_csst_gr}
\end{figure}

\begin{figure}
    \centering
    \includegraphics[width=\linewidth]{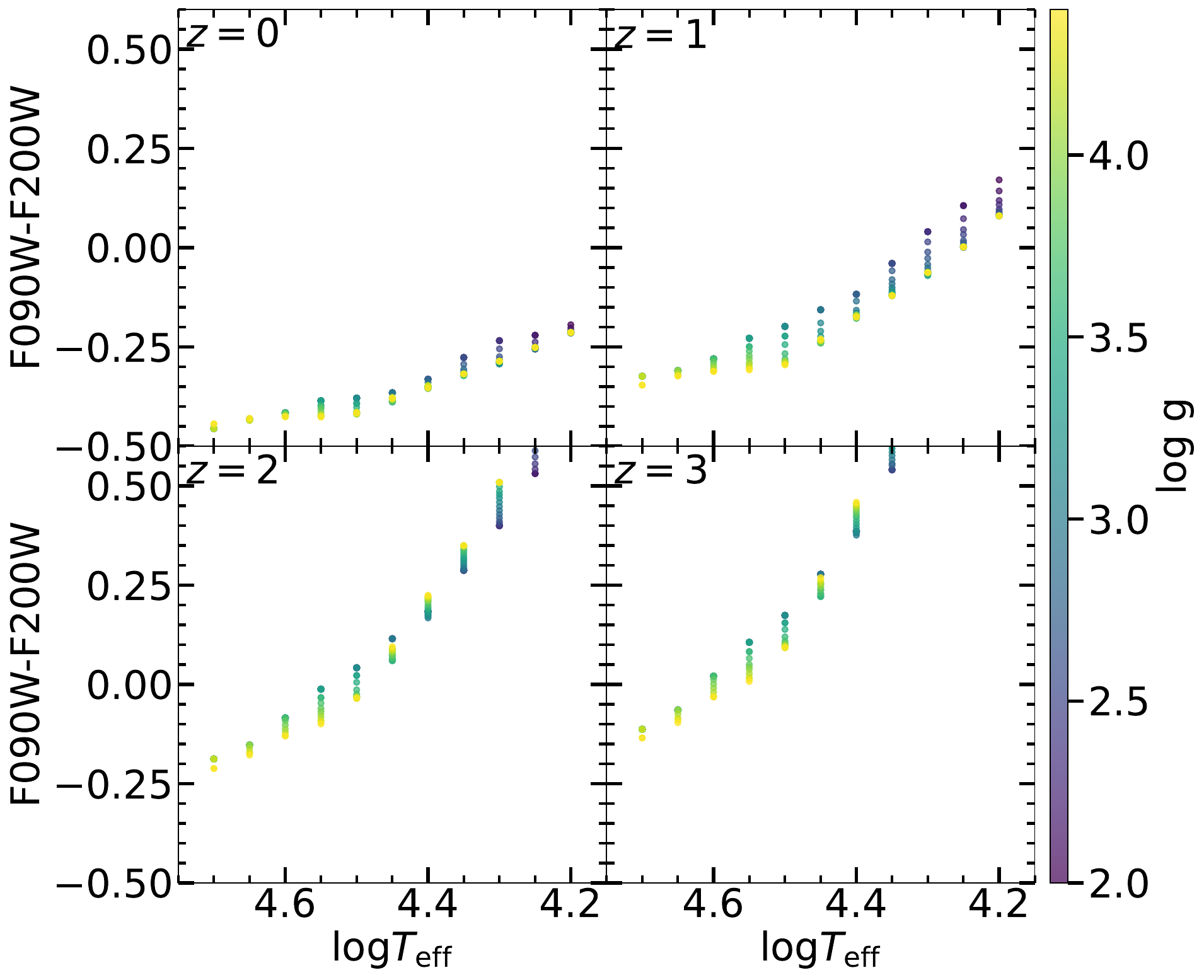}
    \caption{JWST/NIRCam F090W-F200W colour as a function of $\teff$ for the PoWR-OB library.}
    \label{fig:color_teff_PoWROB_jwst}
\end{figure}

To illustrate the BCs computed, we show the CSST/MSC $r-i$ colour as a function of $\logte$ for different redshifts in figure~\ref{fig:color_teff_phoenix_csst_ri}. In this figure, the BCs are computed without any extinction. At redshift zero, the $r-i$ colour follows a very narrow relation with $\logte$ for all the $\logg$ values. The colour dispersion is less than 0.1\,mag in the $\logte$ range shown. In contrast, at high redshifts, the relation shows a much larger dispersion for different $\logg$ values. Such behaviour also occurs for other colours, and for other libraries, as shown in figure~\ref{fig:color_teff_phoenix_csst_gr} for CSST/MSC $g-r$ and figure~\ref{fig:color_teff_PoWROB_jwst} for JWST/NIRCam F090W-F200W. The range of the dispersion changes for the corresponding filter combination and redshifts. We also notice that at high redshifts, the colours are steeper functions of $\logte$. These features indicate that we can have better constraints for $\logg$ and $\logte$ for cosmologically redshifted stars compared to their nearby counterparts, once the redshift can be reliably determined through other methods, such as spectroscopic measurements of the host galaxy. The cause of larger dispersion at higher redshift can be understood by the fact that the shorter wavelength part of the stellar spectra is more sensitive to $\logg$ and gradually shifts to the observed filters. In longer wavelength bands (such as JWST/NIRCam), the dispersions become smaller, which is due to the fact that these bands sample the Rayleigh-Jeans tail and are insensitive to the change of $\logg$. The cause of the steeper relation can be understood with the reddening by redshift.

\begin{figure}
    \centering
    \includegraphics[width=\linewidth]{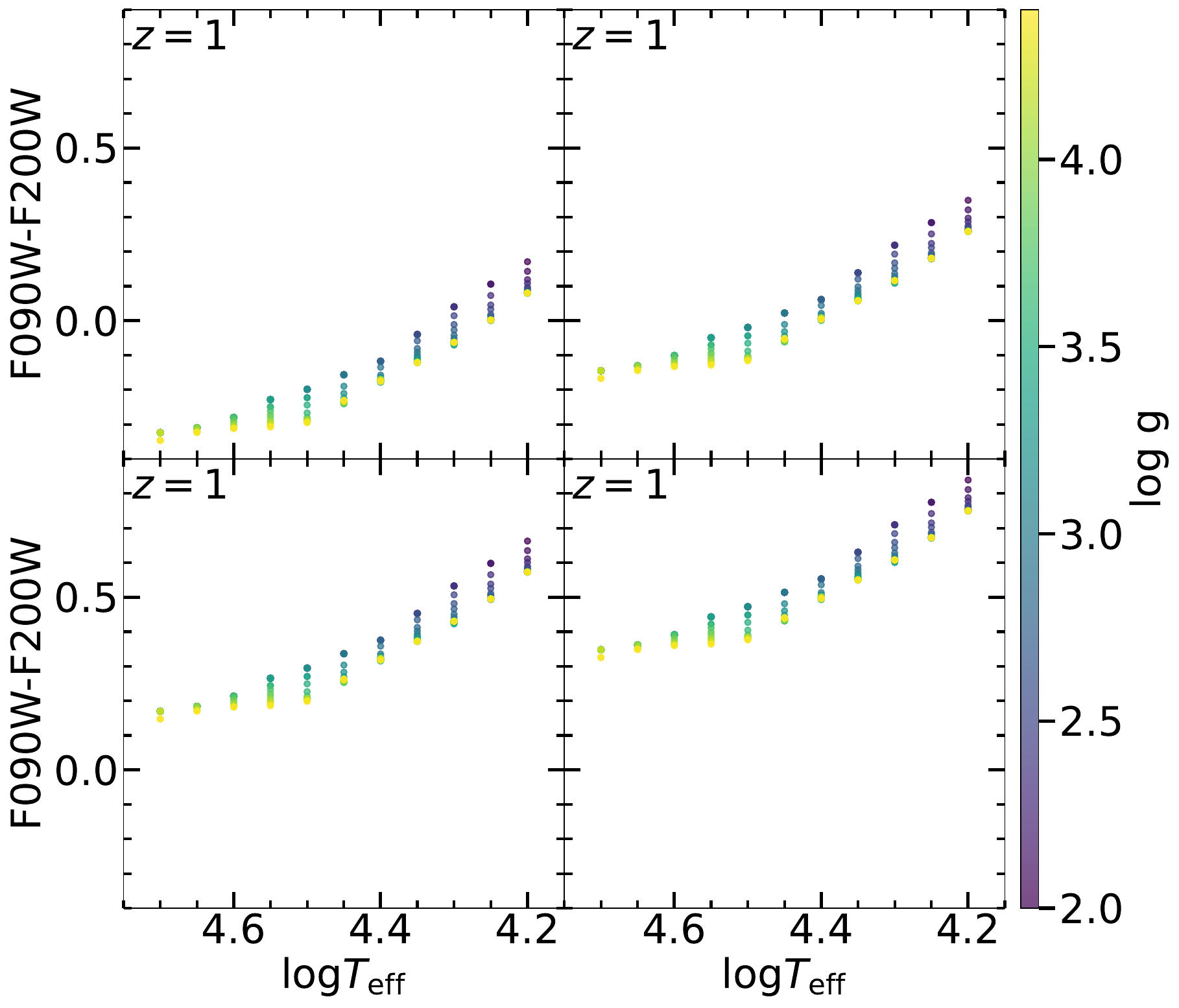}
    \caption{Colours as a function of $\logT$ for the CSST/MSC for the PoWR-OB library with different environmental extinction. Upper left panel: without any extinction, upper right: with Galactic extinction $\av=0.5$, lower left panel: with environmental extinction $\alpha_{V}=0.5$, lower right panel: with Galactic extinction $\av=0.5$ and with environmental extinction $\alpha_{V}=0.5$.}
    \label{fig:color_teff_PoWROB_csst_av}
\end{figure}

To examine the effect brought by the dust, we show the colours as a function of $\logteff$ with different extinctions in figure~\ref{fig:color_teff_PoWROB_csst_av}. In this figure, the nominal values for both the Galactic extinction and environmental extinction are set to 0.5 to evaluate the difference. We see that the effect by environmental extinction is larger than that of the Galactic extinction. We further checked that the difference is even larger for higher redshifts. This can be understood by the fact that the environmental extinction takes effect in the rest-frame of the star, which is in shorter wavelength and has larger extinction coefficients for the same extinction amount. The difference in the extinction curve should also have some minor effects.

\section{CMDs of massive star tracks with redshifts}
\label{sec:track}

\begin{figure*}[ht!]
    \centering
    \includegraphics[width=0.49\linewidth]{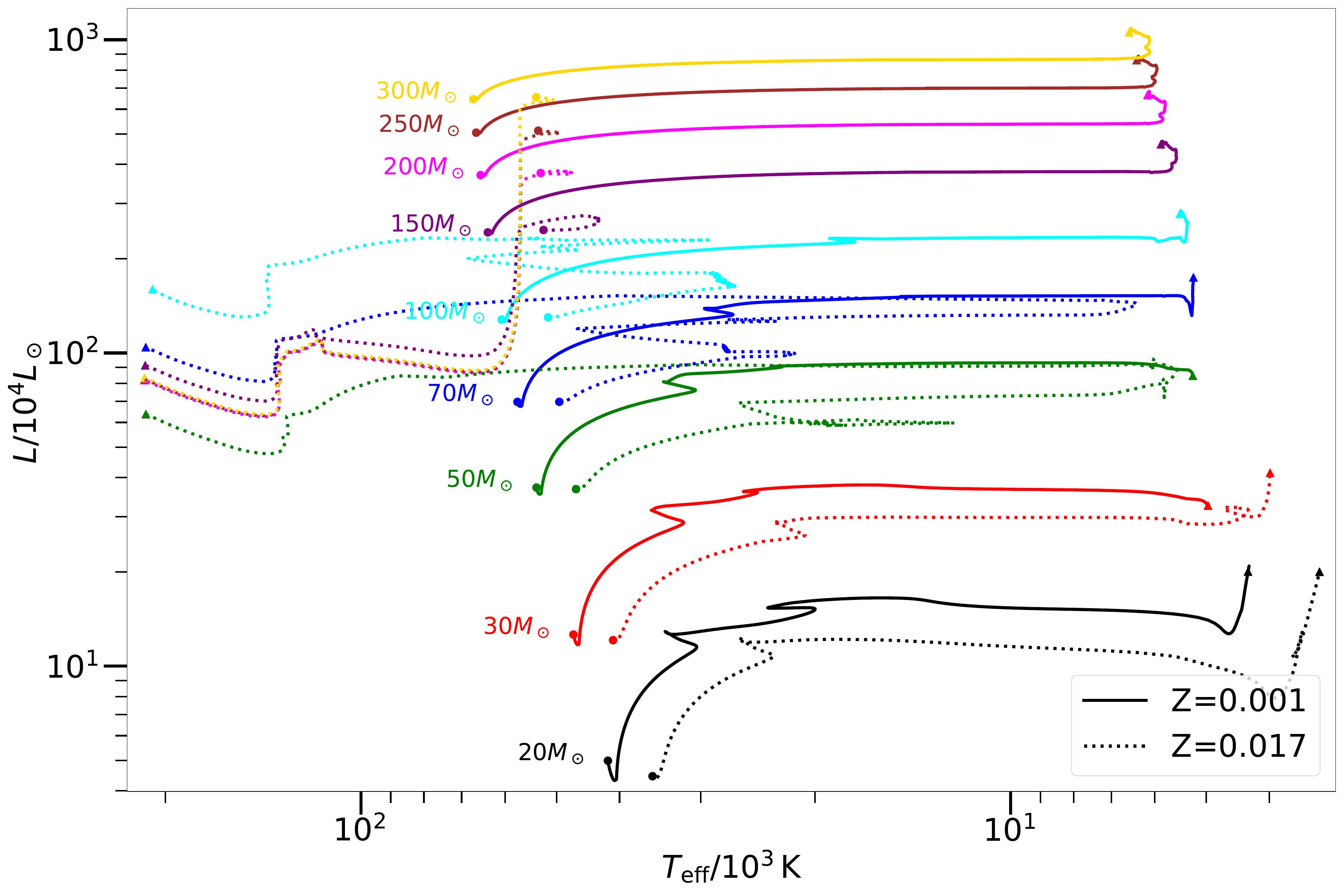}\\
    \includegraphics[width=0.49\linewidth]{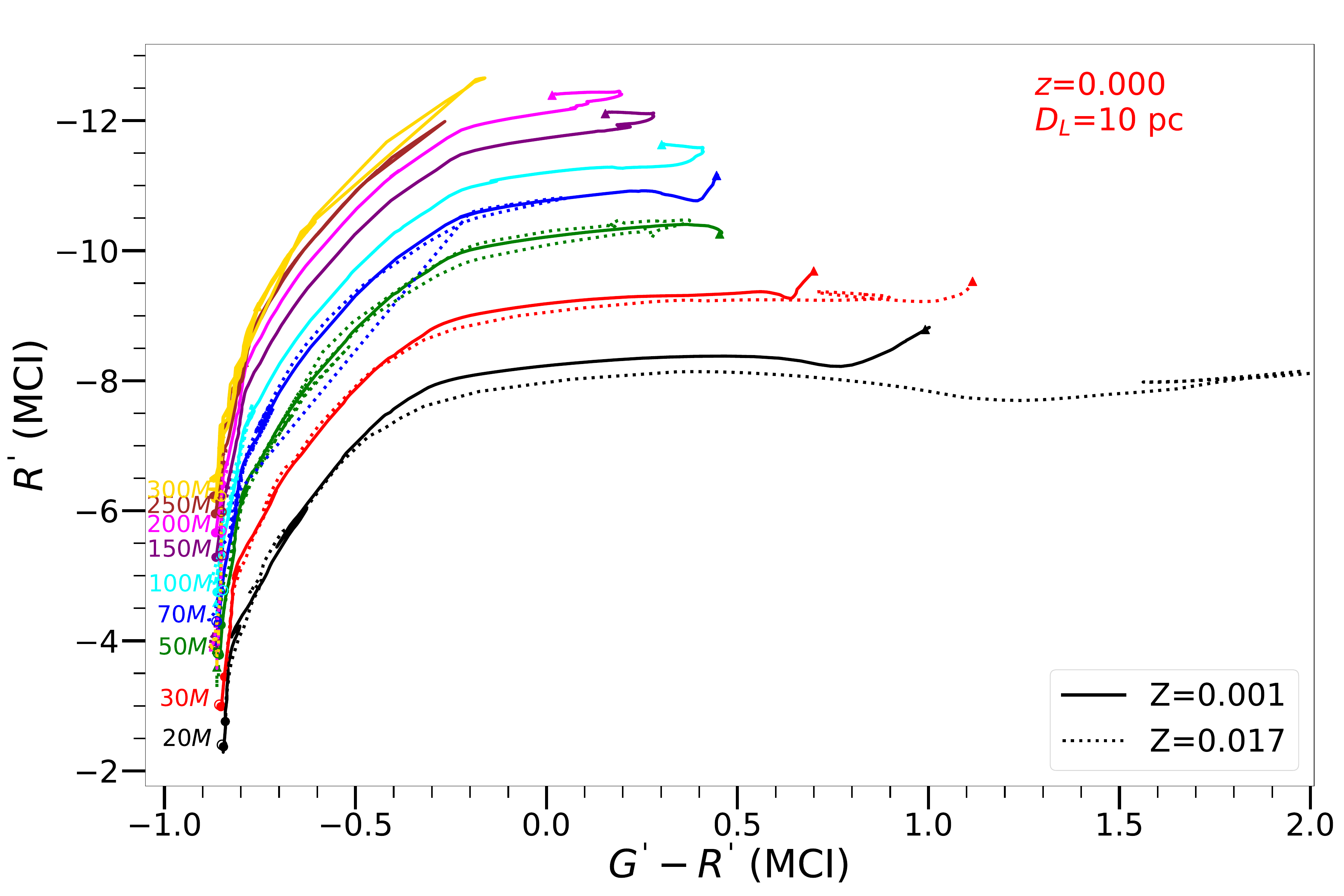}
    \includegraphics[width=0.49\linewidth]{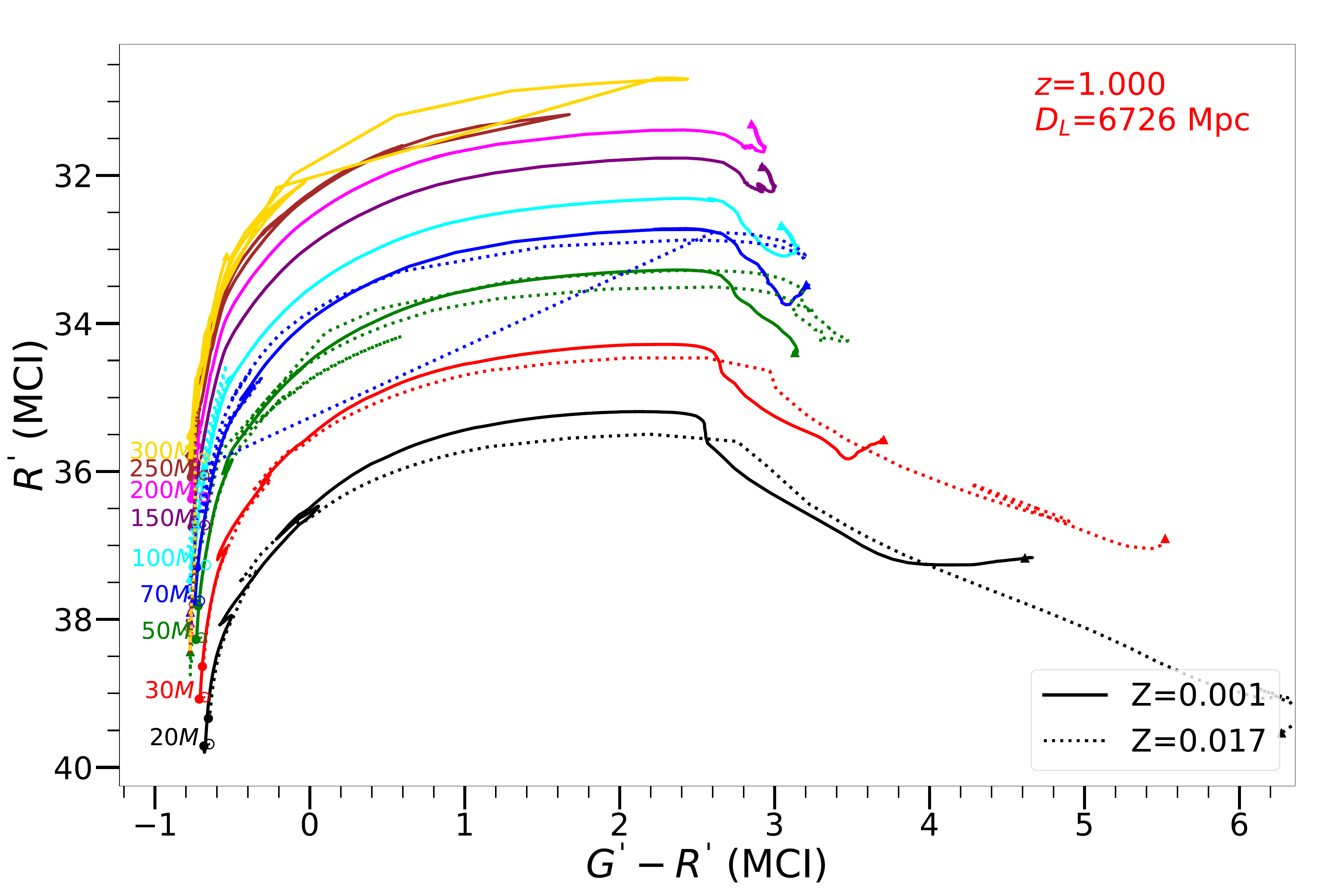}\\
    \includegraphics[width=0.49\linewidth]{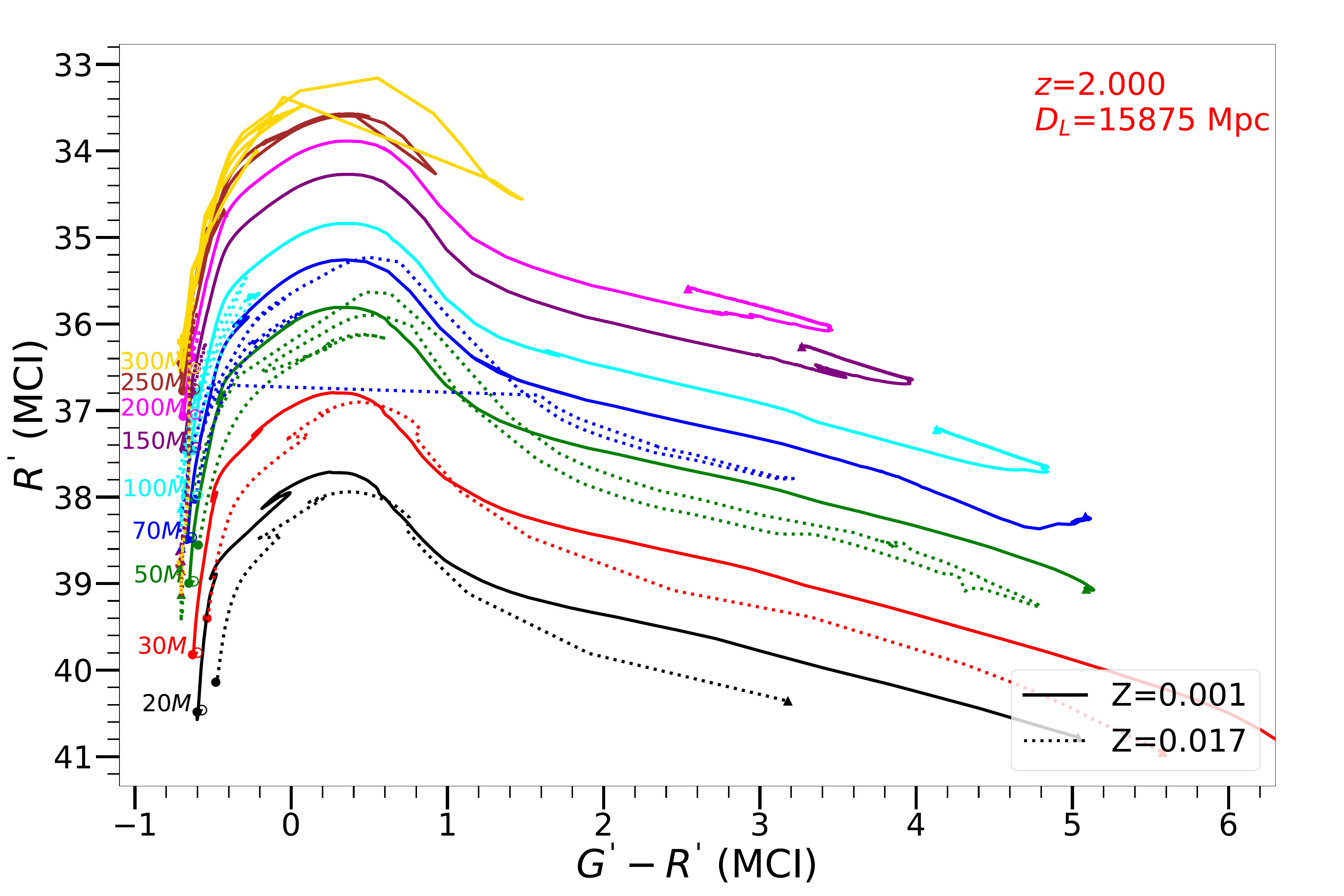}
    \includegraphics[width=0.49\linewidth]{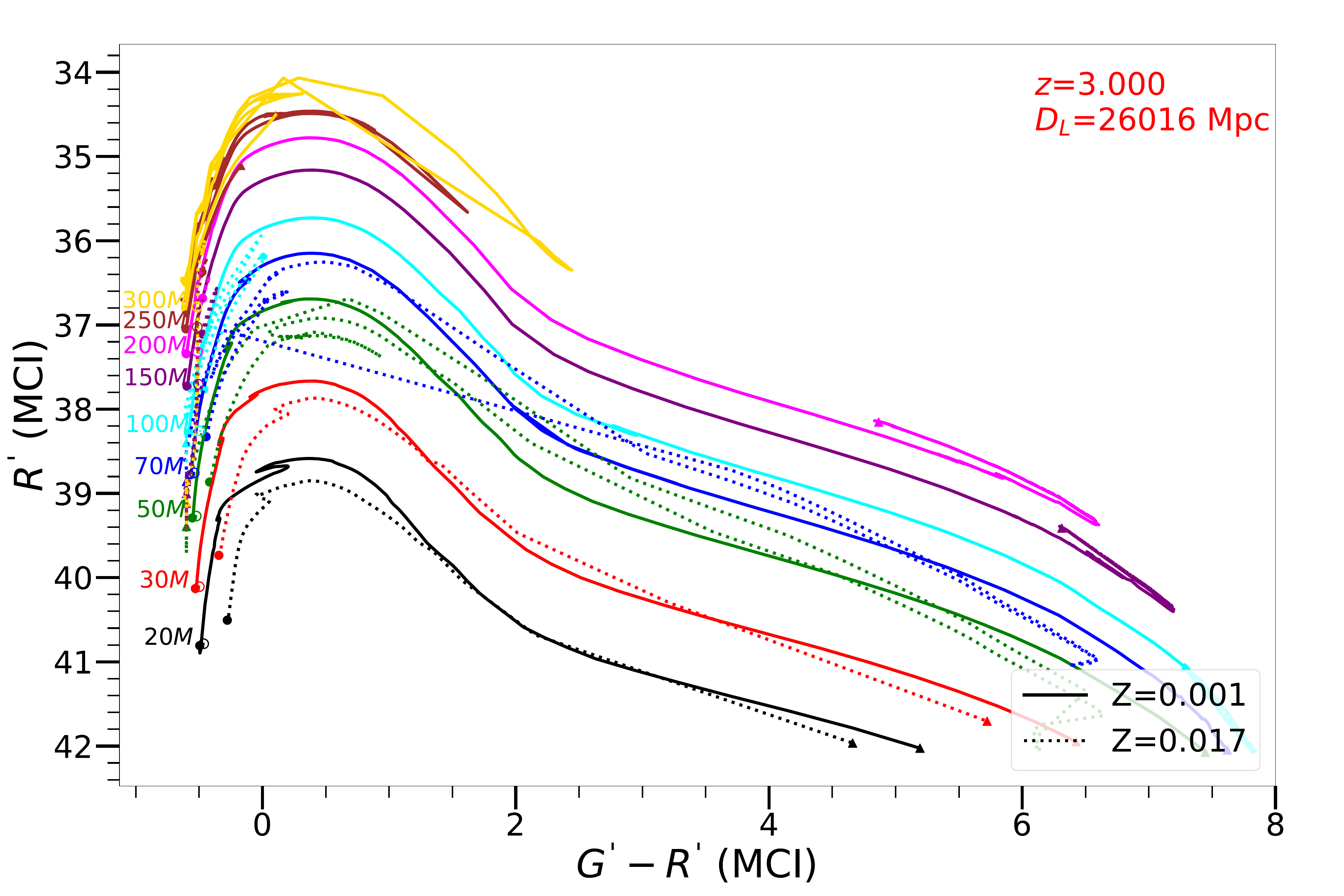}
    \caption{PARSEC evolutionary tracks for massive stars interpolated with the PHOENIX/LYON/BT-Settl set of zYBC database for the CSST/MCI filters. The start points (zero-age main sequence) of the tracks are indicated by the filled dots, while the end points (carbon-ignition) are indicated by the filled triangles.}
    \label{fig:tracks_csst}
\end{figure*}

This database can be useful for the identification and analysis of the properties of cosmologically redshifted star candidates. As a demonstration, we interpolate the PARSEC stellar evolutionary models for massive stars \citep{Chen2015} with the PHOENIX/LYON/BT-Settl set of zYBC tables. The results for the CSST/MCI filters are shown in figure~\ref{fig:tracks_csst}. CSST/MCI has the advantage of targeting the same field simultaneously in three channels and will prove essential for SED-fitting of the high redshift stellar candidates magnified by gravitational micro-lensing. In this figure, we see the overall stretching of the supergiants towards redder and relatively fainter parts of the CMD compared to the main sequence. We also observe that zero-age main-sequence points with nearly identical colours in our sample are separated at high redshifts, which can be very important for unraveling the intrinsic properties of cosmologically redshifted stars from photometric data, as main sequence stars dominate the number fraction of stellar populations.

\section{Summary and Discussion}
\label{sec:discuss}

In this work, we present the zYBC database, which provides bolometric corrections for cosmologically redshifted stars and serves as a major update to our previous YBC stellar bolometric correction database.

These bolometric corrections are derived by convolving telescope transmission curves with redshifted and attenuated stellar spectra from public stellar spectral libraries. The adopted extinction curves include both Galactic dust and dust within the host galaxy, while the effects of the intergalactic medium and intracluster medium are neglected. Our methodology naturally incorporates the cosmological $K$-correction and dust effects. The supported photometric systems include HST/WFC3 wide-band filters, JWST/NIRCam, CSST/MSC, and CSST/MCI, pivotal facilities for current and future observational studies of cosmologically redshifted stars. We also retain the flexibility to incorporate additional photometric systems upon request. In addition to the stellar spectral libraries included in the previous version of YBC, we have added stellar spectral libraries from non-local thermodynamic equilibrium (non-LTE) models, such as PoWR models (for OB stars), TLUSTY, CMFGEN, and updated WM-basic models. These non-LTE models are more suitable for hot massive stars, with PoWR and WM-basic models offering extensive grids covering different mass-loss rates and metallicities. The interpolation across different model sets can be performed directly on the total metallicity Z or [M/H] by adapting the interpolation code described in \cite{Chen2019}. We note that different spectral libraries use different solar abundance scales, which may introduce systematic differences in the derived stellar parameters and BCs. However, for the broad-band applications in this work, such differences are typically within the observational photometric errors for cosmologically redshifted stars.

The database is provided for redshifts $z = $[0, 0.5, 1.0, 1.5, 2.0, 2.5, 3.0, 3.5, 4.0, 4.5, 5.0], Galactic dust with $A_V = $[0, 0.1, 0.2, 0.3, 0.4, 0.5], and environmental dust with $\alpha_V = $[0, 0.5, 1.0, 2.0, 5.0]. Tables for other parameters can be furnished upon request.

As illustrative examples, we present colours as functions of $\log T_{\text{eff}}$ at various redshifts for several photometric systems. These colour-$\log T_{\text{eff}}$ relations exhibit non-monotonic behaviours, highlighting the need for sophisticated modelling. In particular, we find that the relations show greater dispersion at high redshifts compared to the $z=0$ case. This indicates that the stellar parameters of cosmologically redshifted stars can be better constrained than their local counterparts, provided their redshifts are reliably determined through alternative methods (e.g., from their host galaxies) and their photometric data are of sufficient quality for physical parameter derivation via spectral-fitting techniques.

This database is useful for the identification and property analysis of cosmologically redshifted star candidates. We present the colour-magnitude diagram and further observe that zero-age main sequences with similar colours exhibit distinct separations at high redshifts. This finding holds significant implications for unraveling the intrinsic properties of cosmologically redshifted stars from photometric data, as main-sequence stars dominate the number fraction of stellar populations.

Observations from HST and JWST will continue to reveal gravitationally magnified high-redshift star candidates, leading to an increasing demand for accurate stellar bolometric corrections to facilitate comparisons between theoretical stellar evolutionary models and observational data. Thus, we believe this database constitutes a valuable resource for high-redshift star research. Furthermore, CSST/MCI offers significant advantages in discovering and studying extremely magnified stars at redshift $\sim$ 1--3, due to its wide field of view ($\sim$60 square arcminutes), the capability for simultaneous tri-colour observations ensuring the highest possible accuracy in photometric SEDs and deep observation strategy ($\sim$ 30\,mag). In \cite{Zheng2025ApJ...987...94Z}, we presented the first application of the zYBC to simulate the light curves produced by detached binary stars crossing micro-caustics, revealing the diverse light-curve features. These features, particularly the distinct temporal variations in spectral energy distributions, offer diagnostic tools for distinguishing binary systems from single stars and demonstrate the potential of CSST/MCI to capture these variations.

One of the future additions to this database will be the incorporation of spectra for rotating stars from \cite{Girardi2019}, which is essential for accurately modelling the UV spectral features of massive stars with rotation and for making them observable in the optical and NIR bands. This enhancement will further extend the database's utility in interpreting observations of diverse stellar populations at high redshifts.

\section{Data availability}
This updated database is accessible via our website at \url{https://sec.ahu.edu.cn/zYBC} and \url{https://www.sec.center/zYBC}, and accompanied by available code upon request.

\begin{acknowledgements}
We thank the referee for the very helpful suggestions. Y.C. acknowledges the National Natural Science Foundation of China (NSFC) No. 12573030. X.F. thanks the support of the National Natural Science Foundation of China (NSFC) No. 12573040, No. 12533008, and No. 12203100. Y.C. acknowledges the Natural Science Research Project of Anhui Educational Committee No. 2024AH050049, NSFC No. 12003001, the Anhui Project (Z010118169). We acknowledge the China Manned Space Project with No. CMS-CSST-2021-A08. This work is also supported by Chinese Academy of Sciences President's International Fellowship Initiative grant No.2026PVA0172, 2024PD0058.
\end{acknowledgements}

\bibliographystyle{aa.bst}
\bibliography{ref}

\end{document}